\documentclass[
    ,final    ]
  {aipproc}

\layoutstyle{6x9}

\begin{document}

\title{Correlations at RHIC and the Clustering of Color Sources}


\author{E. G. Ferreiro}{
  address={Departamento de F\'{\i}sica de Part\'{\i}culas, Universidad de
Santiago de Compostela, 15782 Santiago de Compostela, Spain}
}

\begin{abstract}
We present our results on transverse momentum fluctuations and
multiplicity fluctuations 
in the framework of the clustering of color sources.
In this approach, elementary color sources -strings- overlap forming clusters,
so the number of effective sources is modified.
These clusters decay into particles with mean transverse momentum that depends on the
number of elementary sources that conform each cluster and the area occupied by the cluster.
We find a non-monotonic dependence of the $p_T$ and multiplicity fluctuations
with the number of participants.
In our approach, the physical mechanism responsible of these fluctuations
is the same: the formation of clusters of strings that introduces correlations between the produced particles.
\end{abstract}

\maketitle

Non-statistical event-by-event fluctuations in relativistic
heavy ion collisions have been proposed as a probe of phase instabilities
near de QCD phase transition.
In a thermodynamical picture of the strongly interacting system formed in
heavy-ion
collisions, the
fluctuations of the mean transverse momentum or mean multiplicity
are related to the fundamental properties of the system,
such the specific heat,
so they may
reveal information about the QCD phase boundary. In particular, a
phase transition in the
evolution
of the system created in relativistic heavy ion collisions may lead to a
divergence of the specific heat
which could be observed as event-by-event fluctuations.
Here I am going to present our results, in the framework of clustering of color
sources, concerning event-by-event $p_T$ and multiplicity
fluctuations.

Event-by-event fluctuations of the transverse momentum
have been measured at SPS
and RHIC energies. The non-statistical fluctuations
show a particular behaviour as a function of the centrality of the
collision: they grow as the centrality increases, achieving a maximum at
mid centralities, followed by a decrease at larger centralities.
Different mechanisms
have been proposed in order to explain those data: complete or partial equilibration,
critical phenomena, as string clustering or string percolation,
and jets production.

Let us concentrate on the results obtained in the framework of clustering of color sources \cite{Ref0}.
In this framework, we consider that in each collision color strings are
stretched between the projectile and the target. Those strings act as the sources of particle production: particles are created via
sea $q-{\bar q}$ production in the field of the string. 
Moreover, in the transverse space, the
color strings correspond to small areas
filled
with the color field created by the colliding partons.

With growing energy and/or atomic number of the colliding
nuclei, the number of sources grows, so the elementary color sources start to
overlap, forming clusters, very much
like disk in the 2-dimensional percolation theory.
The density of strings is expressed by $\eta=N_{st}\frac{S_1}{S_A}$, where
$N_{st}$ corresponds to the total number of strings, $S_1=\pi r_0^2$ with $r_0
=0.2$ fm is the area of each individual string and $S_A$ is the nuclear overlap area.
In particular, at a certain critical density, $\eta_c=1.1\div 1.2$,
a macroscopic
cluster appears, which marks the
percolation phase transition.
Percolation means that a cluster is formed through the whole collision area.

Taking into account that the color charge of a cluster
is the vectorial sum of the
string charges that come into the cluster, 
one can calculate,
for a cluster of $n$ overlapping strings covering an
area $S_n$,
the multiplicity and $p_T$ of
the produced particles :
\begin{equation}
\label{eq1}
Q_n=\sqrt{\frac{nS_n}{S_1}}Q_1\, , \ \ \
\mu_n=\sqrt{\frac{nS_n}{S_1}}\mu_1\, , \ \ \
\langle p_T^2\rangle_n=\sqrt{\frac{nS_1}{S_n}}\langle p_T^2\rangle_1\ .
\end{equation}

In the clustering approach, the behaviour of the transverse momentum
fluctuations can be
understood as
follows:
at low density, most of the particles are
produced by individual strings with
the same $<p_T>_1$, so fluctuations are small.
At large density,
above the critical point, we have only one
cluster, so
fluctuations are not expected either -{\it equilibration}-.
The fluctuations will be maximal
just below the percolation critical density, where there are a large number of
clusters formed by different number of strings
with different size and different $<p_T>_n$.

In orther to measure the event-by-event $p_T$ fluctuations, the proposed
variables are
$F_{p_T}$ and $\phi$, which quantify the deviation of the observed fluctuations from
statistically independent particle emission:
\begin{equation}
\label{eq2}
F_{p_T} = \frac{\omega_{data} - \omega_{random}}{\omega_{random}},\, \ \ \
\omega= \frac{\sqrt{<p_T^2>-<p_T>^2}}{<p_T>},\, \ \ \
\phi=\sqrt{\frac{<Z^2>}{<\mu>}}-\sqrt{<z^2>}\ .
\end{equation}
$z_i={p_T}_i - <p_T>$ is defined for each particle, and $Z_i=\sum_{j=1}^{N_i} z_j$ is defined for each event.

Both variables are related:
$F_{p_T}=\frac{\phi}{\sqrt{<z^2>}}=\frac{1}{\sqrt{<z^2>}}\sqrt{\frac{<Z^2>}{<\mu>}} -1$.
We have computed $F_{p_T}$ \cite{Ref1} using a Monte Carlo code to evaluate
the cluster formation and the analytical expressions (\ref{eq1})
for the
transverse momentum and the multiplicities of the clusters.
\begin{figure}
\resizebox{1.\columnwidth}{!}
{\includegraphics{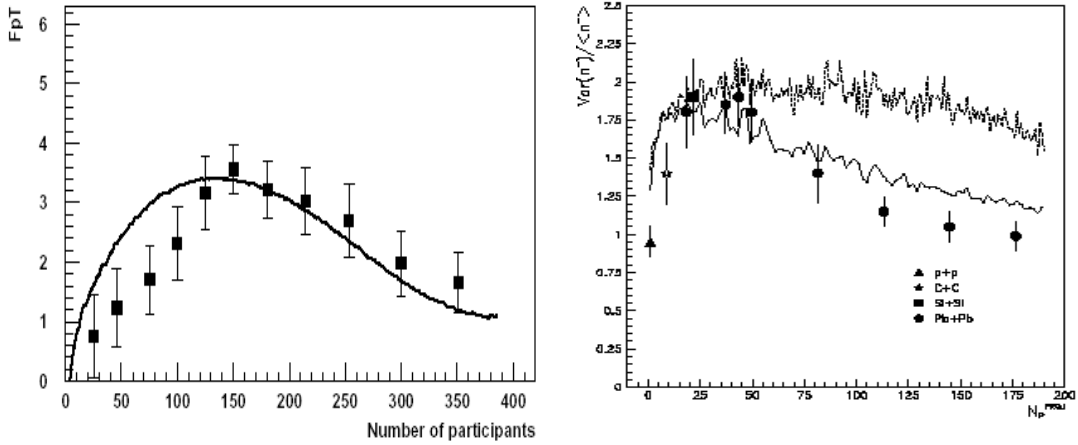}}
  \caption{
Left:
$F_{p_T} (\%)$
versus the number of participants. Experimental
data from PHENIX at $\sqrt{s}=200$ GeV are compared with our results
(solid line). 
Right:
Our results for the scaled variance of negatively charged particles in Pb+Pb collisions at
$P_{lab}=$158 AGeV/c
compared to NA49
experimental data. The dashed line corresponds to our result when clustering formation
is not included, the continuous line takes into account clustering.}
\end{figure}
The behaviour of the transverse momentum
fluctuations
 with the
centrality of the collision shown by the
RHIC
data is naturally explained by
the clustering of color sources.
In this framework,
elementary color sources -{\it strings}- overlap forming
clusters, so the number of effective sources is modified.
These clusters decay into particles with mean transverse momentum that
depends
on the number of elementary sources
 that conform each cluster, and the area
occupied by the cluster.
The transverse momentum fluctuations
 in this approach correspond
to the fluctuations of the transverse momentum
of these clusters,
and they
behave essentially as the number of effective
sources.
In a jet production scenario,
the mean $p_T$ fluctuations are attributed
to jet production in peripheral events, combined with
jet suppression at larger centralities.

A  way to discriminate between the two approaches is to study the
fluctuations at SPS energies \cite{Ref2}, where jet production cannot play a 
fundamental role. Recently, the NA49 Collaboration have presented their data on
multiplicity fluctuations as a
function of centrality at SPS energies. In order to measure these fluctuations, 
the variance of the multiplicity distribution scaled to the mean value of
the multiplicity, 
$Var(N)=\frac{<N^2>-<N>^2}{<N>}$,
has been used.
A non-monotonic centrality -system size- dependence was found. 
In fact, its behaviour is similar to the one obtained for $\Phi (p_T)$
-used by the NA49 Collaboration to quantify the $p_T$-fluctuations-,
suggesting that
they are related to each other.
We find a non-monotonic dependence of the multiplicity fluctuations
with the number of participants.
The centrality behaviour of these fluctuations is very similar to the one
found for the mean $p_T$ fluctuations.
In our approach,
the mechanism responsible for multiplicity and mean $p_T$ fluctuations
is
the formation of clusters of strings
that introduces correlations between the produced particles.
On the other hand,
the mean
$p_T$
fluctuations have been also
attributed
to jet production in peripheral events,
combined with jet suppression in central
events.
However, this hard-scattering
interpretation, based on jet production and jet suppression,
can not be applied to SPS energies, so it does not explain
the non-monotonic behaviour of the mean $p_T$ fluctuations
neither the relation between mean $p_T$ and multiplicity
fluctuations at SPS energy.
Other possible mechanisms are:
combination of strong and electromagnetic interaction,
dipole-dipole interaction and non-extensive thermodynamics.
Still, it is not clear if these fluctuations have
a kinematic or dynamic origin,
but clustering of colour sources remains a good possibility.

%
%
%


\end{document}